\documentclass [12pt,a4paper]{article}
\usepackage{bbm}
\usepackage{amsfonts}
\usepackage{mathrsfs}
\usepackage {amssymb}
\usepackage {amsmath}
\usepackage{amsthm}
\usepackage{latexsym}
\usepackage {amssymb}
\usepackage {amsmath}
\usepackage{color}
\usepackage{blkarray}
\makeatletter
 \@addtoreset{equation}{section}
 
\makeatother

\usepackage{lastpage}
\usepackage{epsfig}

\begin{document}
\begin{center}
\textbf{\Large{Construction of Cyclic and Constacyclic Codes for $b$-symbol Read Channels Meeting\\ the Plotkin-like Bound } }\footnote { The work of M. Yang was supported by the National Natural Science Foundation of China(NSFC) under Grant 61379139, 11526215 and the strategic Priority Research Program of the Chinese Academy of Sciences under Grant XDA06010701. The work of J. Li was supported by the NSFC under Grant 11501156, 61370089 and the Anhui Provincial Natural Science Foundation under Grant 1508085SQA198. The work of K. Feng was supported by the NSFC under Grant 11471178, 11571007 and the Tsinghua National Lab. for Information Science and Technology. \\
M. Yang is with the State Key Laboratory of Information
Security, Institute of Information Engineering, Chinese Academy
of Sciences, Beijing 100093, China(e-mail:yangminghui6688@163.com; ddlin@iie.ac.cn)\\
J. Li is with the School of Mathematics, Hefei University of Technology, Hefei, 230001, China(e-mail: lijin\_0102@126.com )\\
K. Feng is with the Department of Mathematical Sciences, Tsinghua University, Beijing, 100084, China(e-mail: kfeng@math.tsinghua.edu.cn)}

\end{center}

\begin{center}
\small Minghui Yang,  Jin Li,  Keqin Feng
\end{center}


\noindent\textbf{Abstract}-The symbol-pair codes over finite fields have been raised for symbol-pair read channels and motivated by application of high-density data storage technologies [1, 2]. Their generalization is the code for $b$-symbol read channels ($b>2$).
Many MDS codes for $b$-symbol read channels have been constructed which meet the Singleton-like bound ([3, 4, 10] for $b=2$ and [11]
for $b>2$). In this paper we show the Plotkin-like bound and present a construction on irreducible cyclic codes and constacyclic codes meeting the Plotkin-like bound.

\noindent\textbf{keywords}-Symbol-pair codes, $b$-symbol read channel, Plotkin bound, codes for magnetic storage, cyclic codes.

\section{Introduction}

\ \ \ \ Let $\mathbb{F}_q$ be the finite field with $q$ elements. For $n \geq 3$ and $2\leq b\leq n-1$, we define the following
``$b$-symbol read" $\mathbb{F}_q$-linear mapping
$$\pi_b: \mathbb{F}_q^n\rightarrow( \mathbb{F}_q^{b})^{n}$$
$$\textbf{\emph{x}}=(x_0, x_1, \ldots, x_{n-1})\rightarrow \pi_b(\textbf{\emph{x}})=(\textbf{\emph{x}}_{[b]}, \sigma(\textbf{\emph{x}}_{[b]}), \ldots, \sigma^{n-1}(\textbf{\emph{x}}_{[b]}))$$
where $\textbf{\emph{x}}_{[b]}=(x_0, x_1, \ldots, x_{b-1})$ and the ``$b$-symbols"
$$\sigma^{i}(\textbf{\emph{x}}_{[b]})=(x_i, x_{i+1}, \ldots, x_{i+b-1})\ \ \ \ (0\leq i\leq n-1)$$
are the ``cyclic" shiftings of $\textbf{\emph{x}}_{[b]}$, where for $l\geq n, x_l=x_{l-n}$. The $b$-distance between $\textbf{\emph{x}}$ and $\textbf{\emph{y}}\in  \mathbb{F}_q^n$ is defined as the Hamming distance between $\pi_b(\textbf{\emph{x}})$ and $\pi_b(\textbf{\emph{y}})$:
$$d_b(\textbf{\emph{x}}, \textbf{\emph{y}})=d_H(\pi_b(\textbf{\emph{x}}), \pi_b(\textbf{\emph{y}}))=w_H(\pi_b(\textbf{\emph{x}})-\pi_b(\textbf{\emph{y}}))=w_b(\textbf{\emph{x}}-\textbf{\emph{y}}),$$
where for $\textbf{\emph{z}}\in \mathbb{F}_q^n$, $w_b(\textbf{\emph{z}})=w_H(\pi_b(\textbf{\emph{z}}))$ is called the $b$-weight of $\textbf{\emph{z}}$.

A subset $C$ of $\mathbb{F}_q^{n}$ is called a code with parameters $(n, K, d_b)_q$, where $K=|C|\geq 2$ and
$$d_b=d_b(C)=\min\{d_b(c, c'): c, c'\in C, c\neq c'\}.$$
If $C$ is a linear code, namely $C$ is an $\mathbb{F}_q$-subspace of $\mathbb{F}_q^n$, then $K=q^k$, where $k=\dim_{\mathbb{F}_q}C\geq1$ and $d_b=\min\{w_b(c): 0\neq c \in C\}$. For $b=2$, $C$ is called as symbol-pair code which is used for symbol-pair read channels and motivated by the application of high-density data storage technologies [1, 2]. The generalization for $b>2$ [5] can be considered as error-correcting codes over $b$-symbol read channels. In fact, such codes were first introduced even earlier by Levenshtein [6-8] as the sequence reconstruction problem and used in molecular biology and chemistry.

Basic properties of the codes for $b$-symbol read channels have been explored and several bounds have been given to judge the goodness of such codes [1, 2, 5, 9]. Several constructions of such codes have been presented, particularly for MDS codes which meet the Singleton-like bound [3, 4, 10, 11]. In this paper we consider the case where $d_b>\frac{n(q^b-1)}{q^b}$. In the next section we present a new bound for this case as an analogue of usual Plotkin bound [12, Theorem 5.2.4], and construct a series of irreducible cyclic codes and constacyclic codes $(n, K, d_b)_q$ such that their parameters meet the Plotkin-like bound.

\section{A series of codes for $b$-symbol channel meeting the Plotkin-like bound}

\ \ \ \ \ Firstly we show a new bound on codes for  $b$-symbol channel.

\textsl {Lemma 1} (Plotkin-like bound) Let  $b\geq2$ and $C$ be a code for $b$-symbol channel with parameters $(n,K,d_b)_q$. If $d_b>n\theta_b$ where $\theta_b=\frac{q^b-1}{q^b}$, then

\begin{equation*}
  K\leq\frac{d_b}{d_b-n\theta_b}.\tag{1}
\end{equation*}

Moreover, if the code $C$ meets this bound, namely the inequality (1) is an equality then $C$ is an equi-$b$-distance code. Namely, for any distinct codewords $c$ and $c'$ in $C$, $d_b(c,c')=d_b(=\frac{Kn\theta_b}{K-1})$.
\begin{proof} The proof of this Lemma is the same as the proof of [12], Theorem 5.2.4. We duplicate the proof for convenience of readers. Let $C=\{c^{(1)},c^{(2)},$ $\ldots,c^{(K)}\}$ and consider the following $K\times n$ array over $\mathbb{F}_q^b$
\begin{equation*}
A=\left(
  \begin{array}{ccc}
    \pi_b(c^{(1)})    \\
         \vdots   \\
    \pi_b(c^{(K)}) \\
  \end{array}
\right)=\left(
  \begin{array}{ccc}
    c_{[b]}^{(1)}       & \sigma(c_{[b]}^{(1)})      \ \ \cdots     & \sigma^{n-1}(c_{[b]}^{(1)}) \\
     \vdots                         & \vdots        \ \ \ \ \ \ \ \  \ \ \cdots           & \vdots     \\
   c_{[b]}^{(K)}       & \sigma(c_{[b]}^{(K)})      \ \ \cdots     & \sigma^{n-1}(c_{[b]}^{(K)})
  \end{array}
\right)=(a_{j\lambda}), a_{j\lambda}=\sigma^{\lambda}(c_{[b]}^{(j)})\in \mathbb{F}_q^b.
\end{equation*}

Let $\mathbb{F}_q^b=\{v_1,v_2,\ldots,v_{q^b}\}$ and for each $j (1\leq j\leq q^b)$, $m_{j\lambda}$ be the number of $v_j$ appeared in $\lambda$-th column $(0\leq\lambda\leq n-1).$ Then $\sum_{j=1}^{q^b}m_{j\lambda}=K$ for each $\lambda$ and
\begin{equation*}\begin{split}
K(K-1)d_b & \leq \sum_{1\leq i\neq j\leq K}d_b(c^{(i)},c^{(j)})=\sum_{1\leq i\neq j\leq K}d_H(\pi_b(c^{(i)}),\pi_b(c^{(j)}))\\
                                  & = \sum_{\lambda=0}^{n-1} \sum_{j=1}^{q^b}m_{j\lambda}(K-m_{j\lambda})=nK^2-\sum_{\lambda=0}^{n-1}\sum_{j=1}^{q^b}m_{j\lambda}^2\\
                                  & \leq nK^2-q^{-b}\sum_{\lambda=0}^{n-1}(\sum_{j=1}^{q^b}m_{j\lambda})^2\\
                                  & =nK^2-q^{-b}nK^2=n\theta_bK^2.
                                 \end{split}\end{equation*}
Therefore $(K-1)d_b\leq n\theta_bK$ which implies the inequality (1). Moreover, if $(K-1)d_b=n\theta_bK$, then $d_b=\frac{1}{K(K-1)}\sum_{c,c'\in C,c\neq c'}d_b(c,c')$ which implies that $C$ is an equi-b-distance code.
\end{proof}

Now we construct a series of irreducible cyclic codes over $\mathbb{F}_q$ meeting the Plotkin-like bound given by Lemma 1. We fix the following notations.\\

\noindent(A) $p$ is a prime number, $f\geq1, q=p^f$.

\noindent(B) $n\geq2$, $\gcd(n,p)=1$.

\noindent(C) $s$ is the order of $q (\bmod n)$. Namely, $s$ is the smallest positive integer such that

 $q^s\equiv1(\bmod n)$.

\noindent(D) $Q=q^s$, $\mathbb{F}_Q^{\ast}=\langle\gamma\rangle, Q-1=ne, \alpha=\gamma^e$, then the order of $\alpha$ is $n$ and $\mathbb{F}_q(\alpha)=\mathbb{F}_Q$.

\noindent(E) $T_q^{Q}: \mathbb{F}_Q\rightarrow\mathbb{F}_q$ is the trace mapping from $\mathbb{F}_Q$ to $\mathbb{F}_q$. This is an $\mathbb{F}_q$-linear mapping.\\

We consider the following linear code over $\mathbb{F}_q$,
\begin{flushright}
$C=\{c_\beta=(T_q^Q(\beta),T_q^Q(\beta\alpha),\ldots,T_q^Q(\beta\alpha^{n-1}))\in\mathbb{F}_q^{n}: \beta\in\mathbb{F}_Q\}.\ \ \ \  \ \ \  \ \ \ \ \ (2)$
\end{flushright}
This is a cyclic code. The parity-check polynomial $h(x)\in\mathbb{F}_q[x]$ of $C$ is the minimal (irreducible) polynomial of $\alpha^{-1}$ over $\mathbb{F}_q$. From the definition of $s$ we know that $\dim_{\mathbb{F}_q}C=s(=\deg(h(x))$ and $K=|C|=q^s=Q$. The length of $C$ is $n$.

Let $e'=\gcd(e, \frac{Q-1}{q-1})=1.$ In the following result we consider the case of $e'=1$. Since $e|Q-1$ and $Q=q^s$. It is easy to see that $e'=1$ if and only if $e|q-1$ and $\gcd(e,s)=1$.

\textsl {Theorem 1} Let $C$ be the code over  $\mathbb{F}_q$ defined by (2). Assume that $e'=\gcd(e,\frac{Q-1}{q-1})$ is one (namely, $e|q-1$ and $\gcd(e,s)=1).$ Then for each $b (2\leq b\leq s-1),$  the code $C$ has parameters $[n,k,d_b]_q$ with $k=s, d_b=\frac{Q(q^b-1)}{eq^b}$ and meets the Plotkin-like bound given by Lemma 1.
\begin{proof}
Since $C$ is an $\mathbb{F}_q$-linear code,
\begin{equation*}\begin{split}
d_b=d_b(C)& = \min\{w_b(c):0\neq c\in C\}\\
                                  & = \min\{w_b(c_\beta):\beta\in\mathbb{F}_Q^{\ast}\}=\min\{w_H(\pi_b(c_\beta)): \beta\in\mathbb{F}_Q^{\ast}\}
                                 \end{split}\end{equation*}
where
$$\pi_b(c_\beta)=((c_\beta)_{[b]}, \sigma((c_\beta)_{[b]}),\ldots,\sigma^{n-1}((c_\beta)_{[b]}))$$
 and
$$\sigma^{\lambda}((c_{\beta})_{[b]})=(T_q^Q(\beta\alpha^\lambda), T_q^Q(\beta\alpha^{\lambda+1}),\ldots,T_q^Q(\beta\alpha^{\lambda+b-1})\in\mathbb{F}_q^b\ \ (0\leq\lambda\leq n-1).$$
Therefore, for each $\beta\in\mathbb{F}_Q^{\ast}$, $w_H(\pi_b(c_\beta))=n-N_\beta$, where
\begin{equation*}\begin{split}
N_\beta & =\sharp \{\lambda:0\leq\lambda\leq n-1,T_q^Q(\beta\alpha^\lambda)=T_q^Q(\beta\alpha^{\lambda+1})=\cdots=T_q^Q(\beta\alpha^{\lambda+b-1})=0\}\\
                                   & = \sum_{\lambda=0}^{n-1}(\frac{1}{q}\sum_{y_{0}\in\mathbb{F}_q}
                                   \zeta_p^{T_p^q(y_0T_q^Q(\beta\alpha^{\lambda}))}\frac{1}{q}\sum_{y_{1}\in\mathbb{F}_q}
                                   \zeta_p^{T_p^q(y_1T_q^Q(\beta\alpha^{\lambda+1}))}\ldots\frac{1}{q}\sum_{y_{b-1}\in\mathbb{F}_q}
                                   \zeta_p^{T_p^q(y_{b-1}T_q^Q(\beta\alpha^{\lambda+b-1}))})\\
                                  & = \frac{1}{q^b}\sum_{\lambda=0}^{n-1}\sum_{y_0,\ldots,y_{b-1}\in\mathbb{F}_q}
                                  \zeta_p^{T_p^Q[\alpha^{\lambda}\beta(y_0+y_1\alpha+\cdots+y_{b-1}\alpha^{b-1})]}\\
                                  & =\frac{n}{q^b}+\frac{1}{q^b}\sum_{(0,\ldots,0)\neq (y_0,\ldots,y_{b-1})\in\mathbb{F}_q^b}\sum_{x\in D}\zeta_p^{T_p^Q[x\beta(y_0+y_1\alpha+\cdots+y_{b-1}\alpha^{b-1})]},
                                  \end{split}\end{equation*}
where $D=\langle\alpha\rangle=\langle\gamma^e\rangle$ is the subgroup of $\mathbb{F}_Q^{\ast}$ with order $n$. Let $\widehat{\mathbb{F}_Q^{\ast}}$ be the group of the multiplicative characters of $\mathbb{F}_Q$. We have, for $x\in\mathbb{F}_Q^{*}$,
\begin{displaymath}
 \sum_{\substack{\chi\in \widehat{\mathbb{F}_Q^{\ast}}\\\chi(D)=1}}\chi(x)
 = \left\{ \begin{array}{ll}
e, & \textrm{if $x\in D$}\\
0. & \textrm{otherwise}
\end{array} \right.
\end{displaymath}
Therefore
\begin{align*}
N_\beta & =\frac{n}{q^b}+\frac{1}{q^b}\sum_{(0,\ldots,0)\neq (y_0,\ldots,y_{b-1})\in\mathbb{F}_q^b}\sum_{x\in \mathbb{F}_Q^{\ast}}\zeta_p^{T_p^Q[x\beta(y_0+y_1\alpha+\cdots+y_{b-1}\alpha^{b-1})]}\sum_{\substack{\chi\in \widehat{\mathbb{F}_Q^{\ast}}\\\chi(D)=1}}\frac{1}{e}\chi(x)\\
& =\frac{n}{q^b}+\frac{1}{eq^b}\sum_{(0,\ldots,0)\neq (y_0,\ldots,y_{b-1})\in\mathbb{F}_q^b}\sum_{\substack{\chi\in \widehat{\mathbb{F}_Q^{\ast}}\\\chi(D)=1}}\overline{\chi}(\beta(y_0+y_1\alpha+\cdots+y_{b-1}\alpha^{b-1}))G_Q(\chi)\\
& = \frac{n}{q^b}+\frac{1}{eq^b}\sum_{\substack{\chi\in \widehat{\mathbb{F}_Q^{\ast}}\\\chi(D)=1}}\overline{\chi}(\beta)G_Q(\chi)\sum_{(0,\ldots,0)\neq (y_0,\ldots,y_{b-1})\in\mathbb{F}_q^b}\overline{\chi}(y_0+y_1\alpha+\cdots+y_{b-1}\alpha^{b-1})\tag{3}
\end{align*}
where $G_Q(\chi)$ is the Gauss sum over $\mathbb{F}_Q$ defined by
$$G_Q(\chi)=\sum_{x\in\mathbb{F}_Q^{\ast}}\chi(x)\zeta_p^{T^Q_P(x)}$$
and we use the following facts.

(\uppercase\expandafter{\romannumeral 1}) From $\mathbb{F}_q(\alpha)=\mathbb{F}_Q$ and $2\leq b\leq s-1, Q=q^s$ we know that $\{1,\alpha,\ldots,\alpha^{s-1}\}$
is a basis of $\mathbb{F}_Q/\mathbb{F}_q$ and $y_0+y_1\alpha+\cdots+y_{b-1}\alpha^{b-1}\neq 0$ for $(0,\ldots,0)\neq(y_0,\ldots,y_{b-1})\in \mathbb{F}_q^b$.

(\uppercase\expandafter{\romannumeral 2}) For $\delta\in \mathbb{F}_Q^{\ast}$, $\sum_{x\in\mathbb{F}_Q^{\ast}}\chi(x)\zeta_p^{T_p^Q(\delta x)}=\overline{\chi}(\delta)G_Q(\chi).$

Let $H=\{y_0+y_1\alpha+\cdots+y_{b-1}\alpha^{b-1}: (y_0,\ldots,y_{b-1})\in \mathbb{F}_q^b\}$. Then $H$ is an $\mathbb{F}_q-$
subspace of $\mathbb{F}_Q$ with $\dim_{\mathbb{F}_q}H=b$. Moreover, $\mathbb{F}_q^{\ast}\subseteq H^{\ast}=H\backslash \{(0,\ldots,0)\}$ and $H^{\ast}$ is a disjoint union of $\frac{q^b-1}{q-1}$ coset of $\mathbb{F}_q^{\ast}$ in $\mathbb{F}_Q^{\ast}$. Namely,
$$H^{\ast}=\mathbb{F}_q^{\ast}\times S=\{\delta s: \delta\in \mathbb{F}_q^{\ast}, s\in S\}$$
where $S$ is a representative set of such $\frac{q^b-1}{q-1}$ cosets of $\mathbb{F}_q^{\ast}$ in $H^{\ast}$. By (3) we get
\begin{align*}
N_\beta
& =\frac{n}{q^b}+\frac{1}{eq^b}\sum_{\substack{\chi\in \widehat{\mathbb{F}_Q^{\ast}}\\\chi(D)=1}}\overline{\chi}(\beta)G_Q(\chi)\sum_{y\in H^{\ast}}\overline{\chi}(y)\\
& = \frac{n}{q^b}+\frac{1}{eq^b}\sum_{\substack{\chi\in \widehat{\mathbb{F}_Q^{\ast}}\\\chi(D)=1}}\overline{\chi}(\beta)G_Q(\chi)\sum_{s\in S}\overline{\chi}(s)\sum_{\delta\in\mathbb{F}_q^{\ast}}\overline{\chi}(\delta)\ \ \\
& = \frac{n}{q^b}+\frac{q-1}{eq^b}\sum_{\substack{\chi\in \widehat{\mathbb{F}_Q^{\ast}}\\\chi(D)=\chi(\mathbb{F}_q^{\ast})=1}}\overline{\chi}(\beta)G_Q(\chi)\sum_{s\in S}\overline{\chi}(s)  \tag{4}
\end{align*}
Since $D=\langle\gamma^e\rangle$ and $\mathbb{F}_q^{\ast}=\langle\gamma^{\frac{Q-1}{q-1}}\rangle$, we have $D\mathbb{F}_q^{\ast}=\langle\gamma^{e'}\rangle$ where $e'=\gcd(e,\frac{Q-1}{q-1})=1$ by assumption. Therefore the summation in the right-hand side of (4) remains only trivial character $\chi=1$. We get $N_\beta=\frac{n}{q^b}-\frac{q^b-1}{eq^b}=\frac{Q-q^b}{eq^b}$ since $G_Q(\chi)=-1$ for trivial character $\chi=1$, and $|S|=\frac{q^b-1}{q-1}$. Then $w_b(c(\beta))=n-N_{\beta}=\frac{(Q-1)q^b-Q+q^b}{eq^b}=\frac{Q(q^b-1)}{eq^b}$ which is independent of $\beta\in\mathbb{F}_Q^{\ast}$. Therefore $C$ is an equi-b-distance code with $d_b=\frac{Q(q^b-1)}{eq^b}$. From $n=\frac{Q-1}{e}$
and $K=q^k=Q$ we know that $d_b>n\theta_b (\theta_b=\frac{q^b-1}{q^b})$ and $K=d_b/(d_b-n\theta_b)$ which means that the code $C$ for
$b$-symbol channel meets the Plotkin-like bound given by Lemma 1.
\end{proof}

By taking $e=q-1$, the cyclic code $C$ has parameters $[n,k,d_b]_q$ where $n=\frac{Q-1}{q-1}, k=s$ and $d_b=\frac{Q(q^b-1)}{(q-1)q^b}$. In fact, we can get more linear codes by shorten the code $C$ with such parameters. Let $q-1=el$,
then $n=\frac{Q-1}{e}=l\cdot\frac{Q-1}{q-1}$.\\
We consider the following linear code over $\mathbb{F}_q$

\begin{equation*}
\widetilde{C}=\{\tilde{c}(\beta)=(T_q^Q(\beta), T_q^Q(\beta\alpha),\ldots,T_q^Q(\beta\alpha^{\tilde{n}-1}))\in\mathbb{F}_q^{\tilde{n}}: \beta\in\mathbb{F}_Q\}\tag{5}
\end{equation*}
where $\tilde{n}=\frac{Q-1}{q-1}=n/l.$ Namely, we take $\widetilde{c}(\beta)$ as the first $\widetilde{n}$ components of $c(\beta)$. For any $t\geq 1$,
\begin{equation*}
T_q^Q(\beta\alpha^{i+t\tilde{n}})=T_q^Q({\beta\alpha^i})\alpha^{t\tilde{n}}\tag{6}
\end{equation*}
Since $\alpha^{t\tilde{n}}=(\alpha^{\tilde{n}})^t\in\mathbb{F}_q^{\ast}=\langle\gamma^{\tilde{n}}\rangle$. It is easy to see from (6)
that for any $b$ $(2\leq b\leq s-1),$ $w_b(c(\beta))=lw_b(\tilde{c}(\beta))$. Therefore $\tilde{d_b}=d_b(\tilde{C})=d_b(C)/l=\frac{Q(q^b-1)}{(q-1)q^b}$, and $K=\frac{d_b}{d_b-n\theta_b}=\frac{\tilde{d_b}}
{\tilde{d_b}-\tilde{n}\theta_b}.$ We get the following consequence of Theorem 1.

\textsl {Theorem 2} Let $Q=q^s, \mathbb{F}_q^{\ast}=\langle\gamma\rangle, q-1=el, \tilde{n}=\frac{Q-1}{q-1}, \alpha=\gamma^e$ and
assume that $\mathbb{F}_q(\alpha)=\mathbb{F}_Q$ and $\gcd(e, \tilde{n})=1$ (namely, $\gcd(e,s)=1)$. Then for $2\leq b\leq s-1,$ the code $\tilde{C'}$ over $\mathbb{F}_q$ for $b$-symbol channel defined by (5) has parameters $[\tilde{n}, k=s, \tilde{d}_b]_q$ where $\tilde{d_b}=\frac{Q(q^b-1)}{(q-1)q^b}$ and meets the Plotkin-like bound.

\textsl {Remarks} (1) For $b\geq s$, $d_b(C)=n$ and $d_b(\tilde{C})=\tilde{n}$.

(2) The linear code $\tilde{C}$ over $\mathbb{F}_q$ is a ``constacyclic". Namely, if $(c_0, c_1, \ldots, c_{\tilde{n}-1})\in\tilde{C}$
then $(c_1, \ldots, c_{\tilde{n}-1}, \delta c_0)\in\tilde{C}$ where $\delta=\alpha^{\tilde{n}}\in\mathbb{F}_q^{\ast}$.

(3) For the case $e'=\gcd(e, \frac{Q-1}{q-1})\geq 2$, in order to determine $d_b(C)$ we need to compute the summation in the right-hand side of (4) for nontrivial character $\chi$ which might be difficult in general.

\end{document}